\documentclass[aps,amssymb,amsmath, onecolumn, pra, superscriptaddress]{revtex4}
\usepackage{graphicx}
\usepackage{epsfig}
\usepackage{dcolumn}
\usepackage{amsmath}
\usepackage{bm}
\usepackage{bbm}
\usepackage{ulem}
\usepackage[colorlinks, citecolor=red]{hyperref}
\usepackage{color}
\usepackage{soul}
\usepackage{centernot}
\usepackage[up]{subfigure}

\usepackage{epstopdf}

\newcommand{\be}{\begin{equation}}
\newcommand{\ee}{\end{equation}}
\newcommand{\bc}{\begin{center}}
\newcommand{\ec}{\end{center}}
\newcommand{\bea}{\begin{eqnarray}}
\newcommand{\eea}{\end{eqnarray}}
\newcommand{\ba}{\begin{array}}
\newcommand{\ea}{\end{array}}

\def\bra#1{\mathinner{\langle{#1}|}}
\def\ket#1{\mathinner{|{#1}\rangle}}
\def\braket#1{\mathinner{\langle{#1}\rangle}}

\begin{document}
\title{Dirac Cellular Automaton from Split-step Quantum Walk}
\author{Arindam Mallick}
\email{marindam@imsc.res.in}
\author{C. M. Chandrashekar}
\email{chandru@imsc.res.in}
\affiliation{Optics and Quantum Information Group, The Institute of Mathematical
Sciences, C. I. T. Campus, Taramani, Chennai 600113, India}

\begin{abstract}
Simulations of one quantum system by an other has an implication in realization of quantum machine that can imitate any quantum system and solve problems that are not accessible to classical computers. One of the approach to engineer quantum simulations is to discretize the space-time degree of freedom in quantum dynamics and define the quantum cellular automata (QCA), a local unitary update rule on a lattice. Different models of QCA are constructed using set of conditions which are not unique and are not always in implementable configuration on any other system. Dirac Cellular Automata (DCA) is one such model constructed for Dirac Hamiltonian (DH) in free quantum field theory. Here, starting from a split-step discrete-time quantum walk (QW) which is uniquely defined for experimental implementation, we recover the DCA along with all the fine oscillations in position space and bridge the missing connection between DH-DCA-QW. We will present the contribution of the parameters resulting in the fine 
oscillations on the Zitterbewegung frequency and entanglement. The tuneability of the evolution parameters demonstrated in experimental implementation of QW will establish it as an efficient tool to design quantum simulator and approach quantum field theory from principles of quantum information theory.
\end{abstract}

\maketitle

In the relativistic quantum field theories the dynamics are defined using continuum description of space and time degrees of freedom. These continuum description have posed challenges for analytical calculations and remained an hurdle to put the theory on a computer. To overcome these challenges, techniques to discretize the dynamical degree of freedom was developed~\cite{Yuk66, Yam84}. The Dirac equation (DE) describing the relativistic motion of a spin $1/2$ particle is one prominent example where the continuous space and time degrees of freedom has been discretized using different techniques~\cite{Wil74, BMS85, Bia94}. The process of discretization has not followed any unique approach, different techniques like the lattice gauge theory~\cite{Kog79, Rot05} leading to same limit have emerged as discrete theory.  An other discrete evolution model developed to study quantum systems is the quantum version of the cellular automaton~\cite{Neu66}, quantum cellular automaton (QCA)~\cite{Bia94, Mey96, W09}. In 
lattice gauge theory the evolution is described by the unitary operator is the exponential of an Hamiltonian involving the whole system at a same time and in QCA the evolution (update) rule of  the system is described by a local unitary operators each involving few subsystems. The QCA can be regarded as a microscopic mechanism for an emergent quantum fields and as a framework to unify a hypothetical Planck scale with the usual Fermi scale of the high-energy physics~\cite{DP14, Bisio}. The QCA which is not derivable by quantizing classical theory can also be used as a framework for quantum theory of gravity~\cite{Joc95, Ver11}. Different QCA models emerging to Dirac Hamiltonian (DH) for spinor with non-zero mass and massless particles is one prominent example that has been reported~\cite{DP14, Bia94,Bisio} and are referred as Dirac Cellular Automata (DCA) or simply as Dirac Automata (DA).

Though QCA and discrete-time quantum walk (QW)~\cite{Mey96, Ria58, Fey86, Par88, ADZ93} are defined differently for evaluation of quantum field and single particle, respectively at lattice site, the evolution operators for both are unitary, dynamics acts locally and are translationally invariant. The equivalence relation between the class of QW and QCA is also well established~\cite{HKS05} where QW is considered as a single-particle QCA. In a free quantum field theory where the interactions are not taken into consideration, a single particle QCA can mimic the free quantum field. However, when the QCA is developed to describe the dynamics of a specific free quantum field, the standard form of QW evolution operators will not always reproduce the operators corresponding to QCA in the exact form. For example, the DCA~\cite{Bia94,Bisio} cannot be recovered in exact form from the conventional QW operators~\cite{asymp}. Even when the DE which we will use interchangeably with DH is recovered from the 
conventional composition of QW, all the  intricate features observed in DCA are not reproduced. QW has already played an important role in development of efficient quantum algorithms~\cite{Sal12}, to perform different quantum information processing protocols like quantum transport~\cite{PH08, HM09}, quantum memory~\cite{CB15} and to model the dynamics of various quantum systems like energy transfer in photosynthetic systems~\cite{ECR07, MRL08}. Each step of the QW which is discrete in space and time is a composition of a unitary quantum coin operation with variable parameters followed by a coin dependent position shift operator. These evolution protocol can be engineered to suit our applications and can be related to the physical operations in many quantum systems making an experimental implementation a reality~\cite{KFC09, PLM10, SCP10, BFL10}.
Experimental advancements has also further complimented by recent progress in quantum simulations, where one system has been engineered to simulate another demonstrating the precise control over the quantum systems in laboratory. For example, simulation of discretized quantum fields using cold atoms in optical lattices~\cite{LMT10, BMR10, CMP10, SP11, KM11}, coupled cavity arrays~\cite{HBP06, GTC06, ASB07}, trapped ions~\cite{BR12} and photonic systems~\cite{AW12}.

Our understating of conceptual structure of quantum theory and quantification of quantum behavior has improved by many folds with the advances in quantum information theory~\cite{NC00}. These developments has garnered interest in understanding quantum 
field theory and physics in general from the principle of quantum information 
processing, reviving the Feynman~\cite{Fey82} and Wheeler~\cite{Whe90} paradigm of 
physics as information processing. Quantum algorithms which achieves exponential speedup over fastest known classical algorithm to compute relativistic scattering probabilities in a continuum $\phi^4$ theory have been developed~\cite{JLP12}. The continuum $\phi^4$ theory is a simplest interacting quantum field theory which applies to large number of particles at both weak and strong coupling regimes. In this work our focus is to  establish a link between QW and discretized DE in the form of DCA in full generality. In the process of understanding the potential of QW to simulate DE, its dynamics in various approach to continuum limit was explored. The recovery of massless DH from limiting value in the evolution operator~\cite{Mey96, Fre06, CBS10} and the non-zero mass DH by using the rotational invariance property~\cite{Cha13} and by transforming the coordinate system to the null coordinates~\cite{MD12} has been reported. DE in curved space has been recovered from QW in continuum limit by neglecting the higher 
order derivative terms~\cite{MBD13, AFF15}. Further, by rescaling the wavefunction with the coin parameter in the QW, an electromagnetically coupled massive DE has also been obtained~\cite{MBD14}. But, none of these works draw any direct reference to DCA, a discretization of DE. To establish a one-to-one correspondence to DE and QW, we should be able to show that the discretization of DE (or DH) will also lead to operator form identical to the QW evolution operators along with showing the transition of QW to DE in continuum limit. In this direction, one of the recent result shows that the discretization of DE on the quantum lattice Boltzmann falls within the class of QW~\cite{SFP15}. A generic comparative study of QW and DCA was reported highlighting the similarities and the difference in the form of fine oscillations of probability distribution between the two~\cite{asymp}. Here, we show that the split-step QW in place of standard form of QW will reproduce DCA with all the fine oscillations in the 
probability distribution and the effect of these oscillations on the dynamics, Zitterbewegung frequency and entanglement properties. These studies highlight the potential role of using QW in different forms for wide range of studies including, formulation of quantum field theory from the principles of quantum information theory like entanglement properties and simulation of quantum field theory effects like Zitterbewegung oscillations.

In this report, we will first present the description of DCA and QW. Comparing the evolution operators from both the descriptions we will highlight the similarities and differences. In {\bf Results}, starting from one dimensional split-step QW~\cite{kita} which was defined to investigate topological phases and simulate edge states, we will show the complete recovery of the one-dimensional DCA. All the fine oscillations and the entanglement behaviour observed in DCA but not in conventional QW are recovered using split-step QW. We will discuss the consequences leading to these observations and establish a very generic relation between QW-DCA-DH. We will also present the Zitterbewegung oscillations from the parameters that define split-step QW. This will establish QW, which can be designed according to our requirement as an efficient tool to design quantum simulator and approach both, free quantum field theory as well as dynamics in condensed matter systems from the principles of quantum information theory.

 \vskip 0.2in
{\bf Dirac Cellular Automaton~:~}
Cellular automaton is a generalized tool for computation, where both space and time are discrete and state evolution is local that is, the state at position $x$ and time $t$ depends only on the state of neighbouring positions of $x$ including $x$ itself,
at previous time $(t-\tau)$, where $\tau$ is the discrete time step. 
The state update rule acts synchronously at every position called the lattice point where the unit cells of the lattice are all identical with the underlying 
graph being regular~\cite{kari}. The cellular automaton is called quantum, when the state evolution rules are quantum mechanical \cite{GZ88}.
The QCA was first 
introduced in Ref.~\cite{GZ88} and with time, different models of QCA have been developed~\cite{Mey96, Wat95}.
Each QCA model put forward by different set of authors 
have used different set of rules to define them uniquely~\cite{W09}. Therefore, QCA is not uniquely defined like its classical counterpart.
However, QCA model follows 
a general rule of using a set of unitary transition on a lattice of finite-
dimensional quantum systems, on a finite neighbourhood scheme and a finite internal degrees of freedom.

Starting from QCA as a framework, constructing the existing quantum field theories and gravitational theories produced from the Planck 
scale to usual Fermi scale has been explored to understand the theory from quantum information perspective. In one of the approach reported 
recently~\cite{Bisio}, DH has been derived from the QCA by constructing the  evolution operator for a system which is (1) unitary, (2) invariant
under space translation, (3) covariant under parity transformation, (4) covariant under time reversal and (5) has a minimum of two internal degrees of 
freedom (spinor). This QCA evolution which recovers DE is named as DCA and is in the form,
\begin{align}
U_{DA} = \left(\begin{array}{cc}
                ~~\alpha T_-   &   - i \beta \\
                - i \beta    &     ~~\alpha  T_+ \\
                \end{array}\right)
\end{align}
which can be re-written in the form~\cite{asymp},
\begin{align}
\label{DCA}
U_{DA} = \alpha \{ T_- \otimes \ket{\uparrow} \bra{\uparrow} + T_+ \otimes \ket{\downarrow} \bra{\downarrow} \} - i \beta (I \otimes \sigma_x), 
\end{align}
where $\alpha$ corresponds to the hopping strength, $\beta$ corresponds to the mass term, \begin{align}
\ket{\uparrow} = \left(\begin{array}{c}
                        1 \\
                        0 \\
                       \end{array} \right);~~~~~~
                       \ket{\downarrow} = \left(\begin{array}{c}
                        0 \\
                        1 \\
                       \end{array} \right);~~~~\sigma_x = \left(\begin{array}{cc}                             
                               0 & 1\\
                               1 & 0\\ 
                              \end{array} \right), 
\end{align} and
$T_{\pm}$ represents a position shift operator of the form, 
\begin{align}
T_{\pm} = \sum_{x \in a\mathbb{Z}} \ket{x \pm a}\bra{x}
\end{align} with $x$ being the integer multiples of lattice spacing, $a$. The lattice can be considered to be either periodic or infinite, such that, 
\begin{align}
T^{\dagger}_- = T_+ , ~~~ T_- T_+ = T_+ T_- = I  = \sum_{x \in a \mathbb{Z}} \ket{x} \bra{x} = \mbox{Identity in space}.  
 \end{align} 
For an infinite lattice, $x \in \{-\infty,..., -2a, -a, 0, a, 2a,...,+\infty\}$. From the unitarity condition of the operator  $ U_{DA} $  we have,  $|\alpha|^2 + |\beta|^2 = 1,
\mbox{Im}(\alpha^* \beta) = 0 \Rightarrow$ arg($\alpha$) = integer$\times \pi $ + arg($\beta$). So, if we don't worry about the overall phase factor $e^{i [ arg(\beta) ] }$,  
that appears in the $U_{DA}$ operator, we can treat $\alpha$ and $\beta$ as real numbers. For a very larger wavelength compared to Planck length $\approx 1.6162 \times 10^{-35} $ m  and for a mass very much lesser
than the Planck mass $\approx 2.1765 \times 10^{-8} $ kg,  the associated Hamiltonian with this unitary operator in momentum basis, produces Dirac Hamiltonian.
  \begin{align}
   H(k) = \frac{a}{c \tau}\left( \begin{array}{cc}
                  -k c     &   ~m c^2 \\
                 ~~ m c^2   &   k c  \\
                  \end{array}\right)
                 \end{align} 
with the identification $\beta = \frac{m a c}{\hbar}$,  $k$ is a eigenvalue of momentum operator, $\hbar$ is $\frac{1}{2 \pi} \times $ Planck's constant,
 $ m $ is the mass of the associated Dirac particle,   $c$ is the velocity of light in free-medium.
 \vskip 0.2in
{\bf Discrete-time Quantum Walk~:~} 
Quantum walks are broadly classified into two types, discrete and continuous time quantum walks. Here we will focus only on the one dimensional discrete version (QW) where 
the particle which evolves in position space $\{ |x\rangle \}$ has two internal degrees of freedom $|\uparrow\rangle$ and $|\downarrow\rangle$. The state at time $t$ as a linear composition of the internal degrees of freedom can be represented by,
 \begin{align}
        \ket{ \Psi(t)} =  \ket{ \Psi^{\uparrow}(t) } \otimes  \ket{\uparrow} +  \ket{ \Psi^{\downarrow}(t) } \otimes   \ket{\downarrow}
        =  \left(\begin{array}{c} \ket{\Psi^{\uparrow}(t)}  \\     \ket{\Psi^{\downarrow}(t)} \\ \end{array}\right). 
        \end{align}
The $\braket{x|\Psi^{\uparrow (\downarrow)}(t)}  = \Psi^{\uparrow (\downarrow)}(x,t)$ will return the probability amplitude of internal state $\ket{\uparrow} (\ket{\downarrow})$ 
at position $x$. Each step of the QW is defined by a unitary quantum coin operation $C$ on the internal degrees of freedom followed by a position shift operation $S$. 
That is, the state at time $(t+ \tau)$ will be,
\begin{align}
\ket{ \Psi(t + \tau )} = S (I \otimes C)\ket{ \Psi(t)}= U_{QW} \ket{ \Psi(t)} 
  \end{align}
The general form of $C$ is, 
\begin{align}
 C & = C (\xi , \theta, \phi, \delta) = e^{i \xi } e^{-i \theta \sigma_x } e^{ -i \phi \sigma_y } e^{-i \delta \sigma_z}   \nonumber \\ 
  &=e^{i \xi}\left( \begin{array}{cc}
e^{-i\delta} (\cos(\theta) \cos(\phi) - i \sin(\theta) \sin(\phi)) & ~~~ - e^{i \delta} (\cos(\theta) \sin(\phi) + i\sin(\theta) \cos(\phi))\\
e^{-i\delta}(\cos(\theta)\sin (\phi) -i \sin(\theta) \cos(\phi)) & ~~~ ~   e^{i \delta} ( \cos(\theta) \cos(\phi)+ i \sin(\theta) \sin(\phi))\\                                                                                                                                                      \end{array} \right) \nonumber \\ 
 &=e^{i\xi}\left( \begin{array}{cc}
                      ~ F_{\theta, \phi,\delta} &      G_{\theta, \phi,\delta} \\
                       -G^{*}_{\theta, \phi,\delta}  &     F^{*}_{\theta, \phi,\delta} \\
                      \end{array} \right)
\end{align}
where,  $\xi$ is global phase angle,  $2 \theta$, $ 2 \phi$, $ 2 \delta$ are the angles of rotations along $x$, $y$, and $z$ axes respectively, and $\sigma_i$ is the 
$ i $~th component of the Pauli spin matrices $\{ \sigma_x, \sigma_y, \sigma_z \}$, which are generators of SU(2) group.  So, in our internal space the rotational periodicity occurs for rotation angle $\zeta = 4 \pi$ 
instead of $2\pi$ which happens in our spatial rotational case. Here $ \zeta \in \{2\theta, 2\phi, 2\delta\}.$  So, throughout this article we will consider $\theta,\phi,\delta \in  [0, 2 \pi].$ 
The position shift operator $S$ on lattice with spacing $a$ is of the form,
 \begin{align}
S & =  \sum_{x \in a\mathbb{Z}} \ket{x-a}\bra{x} \otimes  \ket{\uparrow}\bra{\uparrow}  + \sum_{x \in a\mathbb{Z}} \ket{x+a} \bra{x} \otimes \ket{\downarrow} \bra{\downarrow}  \nonumber\\
&=  \left(\begin{array}{cc}
       T_{-} & 0 \\
       0 & ~T_{+} \\
      \end{array}\right).
\end{align}

The general form of the evolution operator $U_{QW}$ will therefore be,
\begin{align}
  U_{QW} =    e^{i\xi}\left( \begin{array}{ccc}
                 F_{\theta, \phi,\delta}~ T_{-}   &  &   G_{\theta, \phi,\delta}~T_{-} \\
                 -G^{*}_{\theta, \phi,\delta}~T_{+}    &  &   F^{*}_{\theta, \phi,\delta}~T_{+}
               \end{array}\right). 
\label{ConQW}
\end{align}
The state of the system at position $x$ after one step of walk, at time $(t+\tau)$  will take the form, 
\begin{align}
      \bra{x}\ket{\Psi(t+\tau)} =   \bra{x} U_{QW} \ket{\Psi(t)} = \left( \begin{array}{c}      ~~F_{\theta, \phi, \delta} \Psi^{\uparrow}(x+a,t)    +     G_{\theta, \phi,\delta} \Psi^{\downarrow}(x+a,t) \\
            -G^{*}_{\theta, \phi, \delta} \Psi^{\uparrow}(x-a,t)  +     F^{*}_{\theta, \phi, \delta}\Psi^{\downarrow}(x-a,t) \end{array}\right).
     \end{align}
To elucidate the similarities and difference between the evolution operator for QW and DCA given in equation~(\ref{DCA}), we have to simplify the equation~(\ref{ConQW}). 
By neglecting the global phase term $e^{i \xi}$ and substituting $\phi = \delta = 0$, evolution operator for QW takes the form, \begin{align}
U_{QW} = F_{\theta} \big \{ T_- \otimes \ket{\uparrow}\bra{\uparrow} + T_+ \otimes \ket{\downarrow}\bra{\downarrow} \big \} + G_{\theta} \big \{ T_- \otimes \ket{\uparrow}\bra{\downarrow}) +  T_+ \otimes \ket{\downarrow}\bra{\uparrow} \big \}
\label{QWeq}
 \end{align}
 where
\begin{align}
F_{\theta} =  F_{\theta, 0, 0} = \cos (\theta),~~~~ G_{\theta} = G_{\theta, 0 ,0} = - i \sin (\theta) \nonumber. \end{align}    
Comparing $U_{QW}$(equation~(\ref{QWeq})) with the $U_{DA}$(equation~(\ref{DCA})) we can see that the diagonal elements are identical whereas, the off-diagonal elements in $U_{QW}$ differ with a presence of a shift operator in  place of the spatial identity operator in $U_{DA}$. This difference will remain irrespective 
of the choice of parameters $\theta,~\phi$, and $\delta$. The presence of shift operator in both, diagonal and off-diagonal elements of $U_{QW}$ will always result in a zero probability amplitude  at odd (even) positions after even (odd) number of steps of walk when the initial position $x=0$. In case of $U_{DA}$ the evolution will always return a non-zero probability amplitude at all positions irrespective of even or odd number of steps.

By taking the value of $\theta$ in QW coin operation to tend towards zero,
the off-diagonal terms can be ignored and a massless DH can be recovered. 
These were the first results to establish the connection between the QW and expression for massless DE~\cite{Mey96, Fre06, CBS10}. In order to recover the DE for a non-zero mass particle from the QW evolution operator, the evolution was taken into continuum limit and the coordinate was changed to null coordinate~\cite{MD12}. DH  for a non-zero mass particle was also recovered by taking each step evolution operators to continuous form and by introducing a rotation $e^{-i\theta\sigma_y/2}$~\cite{Cha13}. These continuous approximations suppressed the zero probability in alternate position space which is predominately seen in discrete version. The connection only at limiting value of the coin operator and the need to invoke null coordinates or the rotational invariance to recover DH could not completely resolve the connection between the the QW-DCA-DH. Resolving the difference between the DCA and QW will make QW a suitable method to simulate DH accounting to all intriguing features in the dynamics. This can be 
done by  describing a QW which will evolve with non-zero probability amplitude at all position within the range $x=\pm na$, ($n=$ number of steps). We will discuss this in {\bf Results}.

\vskip 0.3in
\bc
{\bf \Large Results}
\ec
\noindent
{\bf DCA from Split-Step QW~:~}
Here we will present the form of QW, split-step QW which will recover DCA with all the fine oscillations and non-zero probability at all positions between the range of $x=\pm na$. Split-step QW which was first introduced to simulate various 
topological phases~\cite{kita}, and this will establish the split-step QW as a generalization of conventional QW. 

In split-step QW each step of the walk is split into two half-steps, which is composed of two quantum coin operations which in general form will be,
 \begin{align}
C(\theta_1,\phi_1,\delta_1) =\left(\begin{array}{cc}
  ~ F_{\theta_1, \phi_1,\delta_1} &      ~~~G_{\theta_1, \phi_1,\delta_1} \\
  -G^{*}_{\theta_1, \phi_1,\delta_1}  &   ~~~  F^{*}_{\theta_1, \phi_1,\delta_1} \\
                      \end{array}\right), \\
             \nonumber  \\
                         C(\theta_2,\phi_2,\delta_2) =\left(\begin{array}{cc}
  ~ F_{\theta_2, \phi_2,\delta_2} &     ~~~ G_{\theta_2, \phi_2,\delta_2} \\
  -G^{*}_{\theta_2, \phi_2,\delta_2}  &  ~~~   F^{*}_{\theta_2, \phi_2,\delta_2} \\
                      \end{array}\right),
\end{align}
and a two half-shift operators,
\begin{align}
S_- = \left( \begin{array}{cc} 
              T_- & ~~ 0\\
              0  &  ~~ I \\
             \end{array}\right),\\
             \nonumber \\
S_+ = \left( \begin{array}{cc}
              I  & ~~ 0\\
              0  & ~~~ T_+
             \end{array}\right).
\end{align}
The operator $S_{-}$ ($S_{+}$) shifts state $\ket{\uparrow}$ ($\ket{\downarrow}$) to the 
left (right) in position space while leaving the state $\ket{\downarrow}$ ($\ket{\uparrow}$) to remain in same position. 
One complete step of the split-step QW is defined as, 
\begin{align}
  U_{SQW} = S_+ \Big(I \otimes  C(\theta_2, \phi_2 ,\delta_2)\Big) S_- \Big( I \otimes  C(\theta_1, \phi_1, \delta_1) \Big),
  \end{align}
where
\begin{align}
 S_- \Big( I \otimes C(\theta_1, \phi_1, \delta_1)  \Big) =  \left( \begin{array}{ccc}
                 F_{\theta_1, \phi_1,\delta_1}~ T_{-}   &  &   G_{\theta_1, \phi_1,\delta_1}~T_{-} \\
                 -G^{*}_{\theta_1, \phi_1,\delta_1}~I    &  &   F^{*}_{\theta_1, \phi_1,\delta_1}~I
               \end{array}\right)
\end{align}
and,
\begin{align}
S_+ \Big( I \otimes C(\theta_2, \phi_2, \delta_2)   \Big) =  \left( \begin{array}{ccc}
                 F_{\theta_2, \phi_2,\delta_2}~ I   &  &   G_{\theta_2, \phi_2,\delta_2}~I\\
                 -G^{*}_{\theta_2, \phi_2,\delta_2}~T_+    &  &   F^{*}_{\theta_2, \phi_2,\delta_2}~T_+
               \end{array}\right).
\end{align}
Therefore,
\begin{align}
U_{SQW} =  \Big( F_{\theta_2,\phi_2,\delta_2} F_{\theta_1,\phi_1,\delta_1}~I.T_- 
-  G_{\theta_2,\phi_2,\delta_2}G^*_{\theta_1,\phi_1,\delta_1}~I.I \Big) \otimes  \ket{\uparrow}\bra{\uparrow} \nonumber\\
 + \Big(F_{\theta_2,\phi_2,\delta_2} G_{\theta_1,\phi_1,\delta_1}~I.T_-
 + G_{\theta_2,\phi_2,\delta_2}  F^*_{\theta_1,\phi_1,\delta_1}~I.I \Big)\otimes \ket{\uparrow}\bra{\downarrow} \nonumber\\
 + \Big( -G^*_{\theta_2,\phi_2,\delta_2}  F_{\theta_1,\phi_1,\delta_1}~T_+.T_- 
 - F^*_{\theta_2,\phi_2,\delta_2} G^*_{\theta_1,\phi_1,\delta_1}~T_+.I\Big)   \ket{\downarrow}\bra{\uparrow} \nonumber\\
 + \Big( - G^*_{\theta_2,\phi_2,\delta_2} G_{\theta_1,\phi_1,\delta_1}~T_+.T_- 
 + F^*_{\theta_2,\phi_2,\delta_2} F^*_{\theta_1,\phi_1,\delta_1}~T_+.I \Big)  \ket{\downarrow}\bra{\downarrow}
\end{align}
\begin{align}
 = \left( \begin{array}{ccc}
           F_{\theta_2,\phi_2,\delta_2} F_{\theta_1,\phi_1,\delta_1}~T_- -  G_{\theta_2,\phi_2,\delta_2}G^*_{\theta_1,\phi_1,\delta_1}~I   & &     F_{\theta_2,\phi_2,\delta_2} G_{\theta_1,\phi_1,\delta_1}~T_- + G_{\theta_2,\phi_2,\delta_2}  F^*_{\theta_1,\phi_1,\delta_1}~I \\ \\
           -G^*_{\theta_2,\phi_2,\delta_2}  F_{\theta_1,\phi_1,\delta_1}~I - F^*_{\theta_2,\phi_2,\delta_2} G^*_{\theta_1,\phi_1,\delta_1}~T_+      &  &    - G^*_{\theta_2,\phi_2,\delta_2} G_{\theta_1,\phi_1,\delta_1}~I + F^*_{\theta_2,\phi_2,\delta_2} F^*_{\theta_1,\phi_1,\delta_1}~T_+  \\
          \end{array}\right).
         \label{UNSDQW} \end{align}
         
         \begin{figure}[h]
\includegraphics[width=18.0cm]{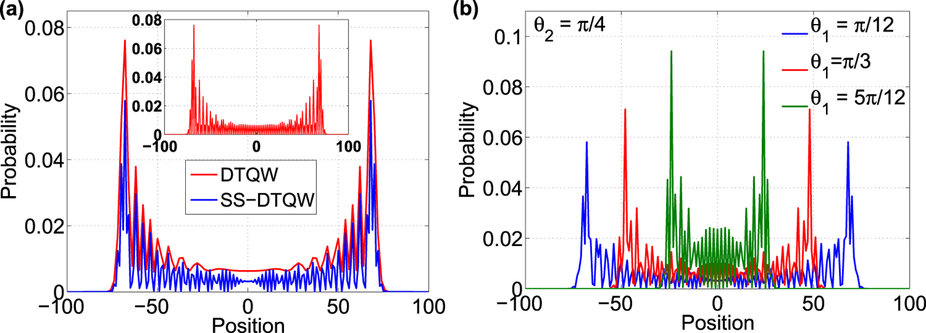}
\caption{{\bf The  probability distribution of finding the particle in one-dimensional position space after 100 steps of conventional and split-step (SS) QW.}  The initial state of the particle and the coin operation used for the evolution are $|\Psi(0)\rangle = \frac{1}{\sqrt 2} (|\uparrow \rangle + |\downarrow \rangle )\otimes |x=0\rangle$ and  $C(\theta_j)$. (a) Blue distribution is for SS-QW ($\theta_1=0$, $\theta_2=\pi/4$) which is identical to DCA when $\alpha = \beta = \frac{1}{\sqrt 2}$ and the red line is for the   conventional QW ($\theta=\pi/4$). Points with zero probability is removed from the main plot whereas, it is retained in the inset. (b) SS-QW for evolution using different combinations of $\theta_1$ and $\theta_2$. From these distribution we can say that the oscillations in the probability distribution is not unique to the combination of $\theta_{j}$ which results in recovery of DCA.}
\label{fig1}
\end{figure}
          In the preceding expression for $U_{SQW}$ we get both, position shift 
          operator and spatial identity operators in the diagonal as well as off-diagonal elements.
          To obtain $U_{SQW}$ in the same form as $U_{DA}$ (equation~(\ref{DCA})), we have to identify the parameters
          of the quantum coin operators that will remove the spatial identify component along the diagonal and spatial-shift 
          component along the off-diagonal elements.
Therefore, the coin parameters should satisfy, 
\begin{align}
G_{\theta_2,\phi_2,\delta_2}G^*_{\theta_1,\phi_1,\delta_1} &= e^{i (\delta_{2}-\delta_{1})} \big[\cos(\theta_{2}) \sin(\phi_{2}) + i\sin(\theta_{2}) \cos(\phi_{2})\big] \big[\cos(\theta_{1}) \sin(\phi_{1}) - i\sin(\theta_{1}) \cos(\phi_{1})\big]= 0\\
F_{\theta_2,\phi_2,\delta_2} G_{\theta_1,\phi_1,\delta_1} &=
e^{i(\delta_1 - \delta_2)}\big [\cos(\theta_{2}) \cos(\phi_{2}) - i \sin(\theta_{2}) \sin(\phi_{2})\big] \big[\cos(\theta_{1}) \sin(\phi_{1}) + i\sin(\theta_{1}) \cos(\phi_{1})\big] = 0.
\end{align}

Among the possible solutions we will choose the parameter $\theta_{1} = \phi_{1}  =\delta_1 = \delta_2 = 0$ which will recover the DCA and satisfy the above conditions. By substituting them in $U_{SQW}$ we get, 
\begin{align}
        U_{SQW} = \left( \begin{array}{cc}
              ~~~\cos(\theta_2) T_{-}  &          -i \sin(\theta_2) I \\
        -i \sin(\theta_2) I    &   ~~~~ \cos(\theta_2) T_{+}
        \end{array}\right),
       \label{coinSQW}
        \end{align}
        which is in the same form as $U_{DA}$ where $\beta = \sin(\theta_2)  \equiv \frac{mc a}{\hbar}$ and $\alpha = \cos(\theta_2)$. 
        From this unitary operator we will recover the Hamiltonian in the form,
\begin{align}
    H_{SQW} = - \frac{ \hbar \cos^{-1} \Big( \cos (\theta_2) \cos \big(\frac{ka}{\hbar}\big) \Big)}{ \tau \sqrt{1 - (\cos (\theta_2) \cos\big(\frac{ka}{\hbar}\big))^2}} \Bigg[ \cos(\theta_2) \sin\Big(\frac{ka}{\hbar}\Big) \left( \begin{array}{cc}
                                                                                                                       1 & 0\\
                                                                                                                       0 & -1 \\
                                                                                                                      \end{array}\right) -\sin (\theta_2) \left( \begin{array}{cc}
                                                                                                                                                               0 & 1\\
                                                                                                                                                               1 & 0 \\
                                                                                                                                                              \end{array}\right)
 \Bigg].
   \end{align}
See {\bf Methods} for the derivation. For smaller mass,  $\theta_2  \approx 0$  and for smaller momentum,  $ k \approx 0,$ $\sin\theta_2 \approx \theta_2, \cos\theta_2 \approx 1, \sin\Big(\frac{ka}{\hbar}\Big) \approx \frac{ka}{\hbar}, \cos \Big(\frac{ka}{\hbar}\Big) \approx 1. $
Then \begin{align}
      H_{SQW} \approx - \frac{a}{\tau}  k \left(\begin{array}{cc}
                                1 & 0\\
                                0 & -1\\
                               \end{array}\right) + \frac{\hbar}{\tau} \theta_2 \left( \begin{array}{cc}
                                                                 0 & 1\\
                                                                 1 & 0\\
                                                                 \end{array}\right)
                               \end{align} which is in a form of one-dimensional DH for a $\frac{1}{2} $ spinor, with the identifications, $\frac{a}{\tau} = c $ 
                               and $ \frac{\hbar \theta_2}{\tau} = m c^2 $, so, $ m = \frac{\hbar \theta_2 \tau}{a^2}$. 

In this section, starting from split-step QW we obtained the expression for DCA and from that we recovered DH without invoking any invariance property explicitly. In Fig.\ref{fig1}(a) we present the probability distribution of QW and split-step QW (same for DCA) after 100 steps of walk using the coin operation of the form $C(\theta_j) =\left(\begin{array}{cc}  ~~\cos(\theta_j)  &  -i\sin(\theta_j)\\
                                -i\sin(\theta_j) & ~~\cos(\theta_j)
                               \end{array}\right)$. Though both the distribution spread along the same envelop, and fine oscillations which is seen in the split-step QW is absent in conventional QW. In the inset we show the probability distribution for conventional QW without removing the points with zero probability at alternate position space. In Fig.~\ref{fig1}(b) we have presented the probability distribution of the split-step QW after 100 steps of evolution using  different combinations of $\theta_{1}$ and $\theta_{2}$. In spite of having the similar probability distributions these combinations do not straight away recover the DCA like it does for the $(\theta_1=0, \theta_2 = \pi/4)$. All the plots in this report were obtained by time iteration evolution of the walk operators.

 \vskip 0.2in
 
\noindent
{\bf Zitterbewegung Oscillation~:~} 
Any quantum mechanical observable $\hat{A}$ which doesn't commute with the Hamiltonian operator,
that is, $[\hat{A}, H ] \neq 0, $ results in mixing of positive and negative energy eigenvalue solutions during the evolution.
This mixing is responsible for oscillation of the expectation value of the observable and is known as Zitterbewegung oscillation~\cite{zit}. Zitterbewegung oscillation is a very common phenomenon that describes the jittering motion of free relativistic Dirac particles,
as predicted by evolution driven by DH. Therefore, we will look into this interesting phenomenon as a function of split-step QW evolution parameter and compare it with the configuration of the parameter for which we see the equivalence with DCA. 

We will first consider the split-step QW with the non-zero parameters $\theta_{1}$ and $\theta_{2}$, with $\phi$'s and $\delta$'s set to zero. From this we can deduce to 
the parameter configuration which results in equivalence with DCA. The evolution parameter as a function of $\theta_{1}$ and $\theta_{2}$ is,
\begin{align}
  \label{SSQW}
  U_{SQW} =  \left( \begin{array}{cc}
              -\sin(\theta_1) \sin(\theta_2) I      &      ~~ ~~   -i \cos(\theta_1) \sin(\theta_2) I \\
         ~     + \cos(\theta_1) \cos(\theta_2) T_{-}   &      ~~  ~~~  - i \sin(\theta_1) \cos(\theta_2) T_{-}\\ \\
             -i \cos(\theta_1) \sin(\theta_2) I     &      ~~~    - \sin(\theta_1) \sin(\theta_2) I  \\
      ~      - i \sin(\theta_1) \cos(\theta_2) T_{+}   &   ~~~~~        + \cos(\theta_1) \cos(\theta_2) T_{+}\\
        \end{array}\right). 
\end{align}

The internal states of the system remain same by the action of the diagonal terms of operator of equation~(\ref{SSQW}),
but the off-diagonal terms flip among $\ket{\uparrow}$ and $\ket{\downarrow}$. The overall effect of the diagonal
terms are simple forward (positive $x$ direction) or backward (negative $x$ direction) movement for individual internal degrees of freedom. But the off-diagonal terms which cause both, flipping in internal degrees and spatial shift, these cause oscillatory movement in $x-$position axis. 
       \begin{figure}[h]
\includegraphics[width=18.0cm]{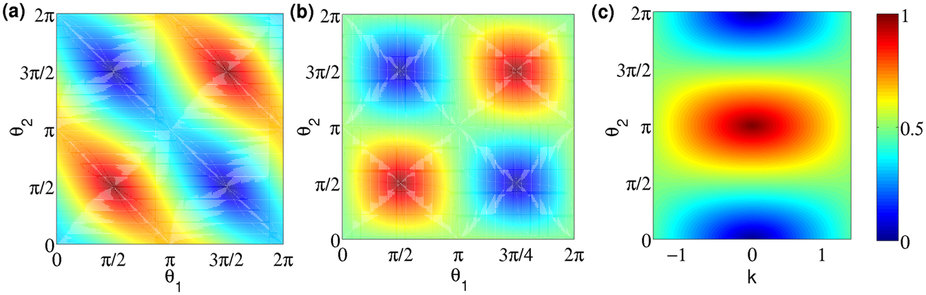} 
\caption{(color online) {\bf Zitterbewegung frequency as a function of $\theta_1$, $\theta_2$ and $k$.} Oscillation frequency as function of $\theta_1$ and $\theta_2$ when (a) $ka/\hbar =1$ and (b) $ka/\hbar = 10^{34}$. (c) Oscillation frequency as function of $k$ (in range $ \{-\sqrt{2}$, $+\sqrt{2} \}$) and $\theta_2$ when $\theta_1 =0$ and $a/\hbar=1$ (SS-QW equivalent to DCA).}
\label{zitter}
\end{figure}

For split-step QW the Zitterbewegung frequency is,                                
               \begin{align}
           Z_{SQW} =  \frac{ 1 }{\tau \pi} \cos^{-1} \Big[ \cos (\theta_1) \cos (\theta_2) \cos\Big(\frac{ka}{\hbar}\Big) - \sin (\theta_1) \sin (\theta_2) \Big]
            \end{align}
        and corresponding amplitude of oscillation,
        \begin{align}
        \label{zitterAMP}
         \mathcal{A}_{SQW} = 2 \sqrt{  \big[\Re(c^*_1 c_2 \bra{\phi^+_k(t=0)} \hat{A} \ket{\phi^-_k(t=0)})\big]^2 +
         \big[ \Im(c^*_1 c_2 \bra{\phi^+_k(t=0)} \hat{A} \ket{\phi^-_k(t=0)})\big]^2}.
       \end{align}
 See {\bf Methods} for the intermediate steps and the extended forms of the terms in equation~(\ref{zitterAMP}).    
                                
                                For case with one-to-one correspondence with DCA, $Z_{SQW}$ reduces to,   
\begin{align}
Z_{SQW} (\theta_1 = 0) \equiv Z_{DCA} = \frac{1}{ \tau \pi} \cos^{-1} \Big[ \cos (\theta_2) \cos\Big(\frac{ka}{\hbar}\Big) \Big].
\end{align}

In Fig.~\ref{zitter}(a) and (b), the Zitterbewegung frequency as a function of $\theta_1$ and $\theta_2$ for two different values of $ka/\hbar$ is shown. The maximum and minimum oscillation frequency is for non-zero $\theta_1$. For configuration leading to DCA, $\theta_1=0$, the oscillations frequency as function of $k$ ranging from $-\sqrt{2}$ to $\sqrt{2}$~\cite{Cha12}, and $\theta_2$ is shown in Fig.~\ref{zitter}(c). With the combination of coin parameters and $k$ one can demonstrate complete control on the frequency of the Zitterbewegung frequency.

\vskip 0.2in

  \noindent
    {\bf Entanglement between position space and internal degree~:~}
QW gives easy access to study the entanglement behavior of the evolving 
particle with the position space. In Ref.~\cite{asymp} it was shown that the entanglement between the particle and the position space for DCA is higher compared to the conventional QW. Since we have show that the split-step QW with $\theta_1=0$ and $\theta_2=\pi/4$ is equivalent to DCA with $\alpha = \beta =\frac{1}{\sqrt 2}$, comparing the entanglement between the split-step QW with conventional QW will suffice to compliment and present the more general observations. 
\begin{figure}[h]
\includegraphics[width=18.0cm]{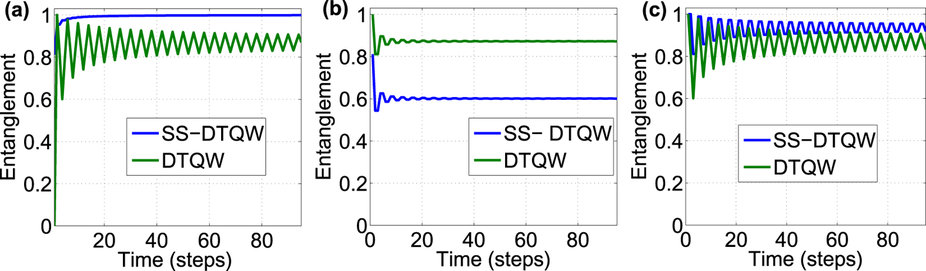}
\caption{(color online) {\bf Entanglement as a function of time with different initial state.} For conventional QW coin parameter $\theta=\frac{\pi}{4}$ and for SS-QW $\theta_1 = 0, \theta_2 = \frac{\pi}{4} $. The initial states in (a) $\frac{1}{\sqrt{2}} (\ket{\uparrow} + i \ket{\downarrow}) \otimes \ket{x=0} $ 
(b) $\frac{1}{\sqrt{2}} (\ket{\uparrow} +  \ket{\downarrow}) \otimes \ket{x=0} $ and 
(c) $ \ket{\uparrow} \otimes \ket{x=0} $. Dependency of entanglement value on the initial state is higher for split-step QW compared to the conventional QW.}
 \label{fig3}
\end{figure}
We will define the initial state in density matrix form on the total Hilbert space $\mathcal{H} = \mathcal{H}_{coin} \otimes \mathcal{H}_x$,
\begin{align}
\rho(0) = \left(\begin{array}{cc}
\cos^2 \frac{\Omega_{p}}{2}  & \sin \frac{\Omega_p}{2} \cos \frac{\Omega_p}{2} e^{-i \Omega_a} \\
&    \\
\sin \frac{\Omega_p}{2} \cos \frac{\Omega_p}{2} e^{i \Omega_a} &  \sin^2 \frac{\Omega_p}{2}  \\
\end{array} \right)\otimes \ket{x=0}\bra{x=0}.
\end{align} is a pure state.
Here, $ \ket{x} \in \mathcal{H}_x, $ $\Omega_p \in [0, \pi] $ and $\Omega_a \in [0, 2 \pi) $ are respectively,
the polar and azimuthal angle of Bloch sphere associated with the coin space.
        The state after time $t$ will be, 
       \begin{align}\rho(t) = U_{SQW}^\frac{t}{\tau} \rho(0) ( U^{\dagger}_{SQW})^\frac{t}{\tau},
       \end{align} 
 where $U_{SQW}$ is given by equation~(\ref{SSQW}).
 
 As here we are dealing with only the evolution of a pure quantum state which remains pure by 
 unitary evolution, we will use the partial entropy as a measure of entanglement, which is enough to give correct measure 
 of entanglement of a pure state.
 For that we first take partial trace with respect to  $\mathcal{H}_x$-space (position space) of time evolved state 
       = $Tr_x(\rho(t)) := \rho_c(t).$
       Then according our measure the entanglement at time $t$ is given by, 
       \begin{align}
       -Tr_c[\rho_c(t) \log_2 \{\rho_c(t)\}],
       \end{align}
the suffix {\it c} represents the coin space. 
In Fig.~\ref{fig3}, we present the value of entanglement as a function of time for conventional QW and split-step QW which recover DCA
for evolution with three different initial states. For the three initial state presented, the mean value of the entanglement remains same for conventional QW with only a change in the fluctuations around the mean value. For the split-step QW, the entanglement itself varies significantly reaching maximum value (one) with the change in the initial state. This variation in entanglement value is an indication of greater change in the degree of interference in split-step QW compared to convention QW.
\begin{figure}[h]
\includegraphics[width=17.0cm]{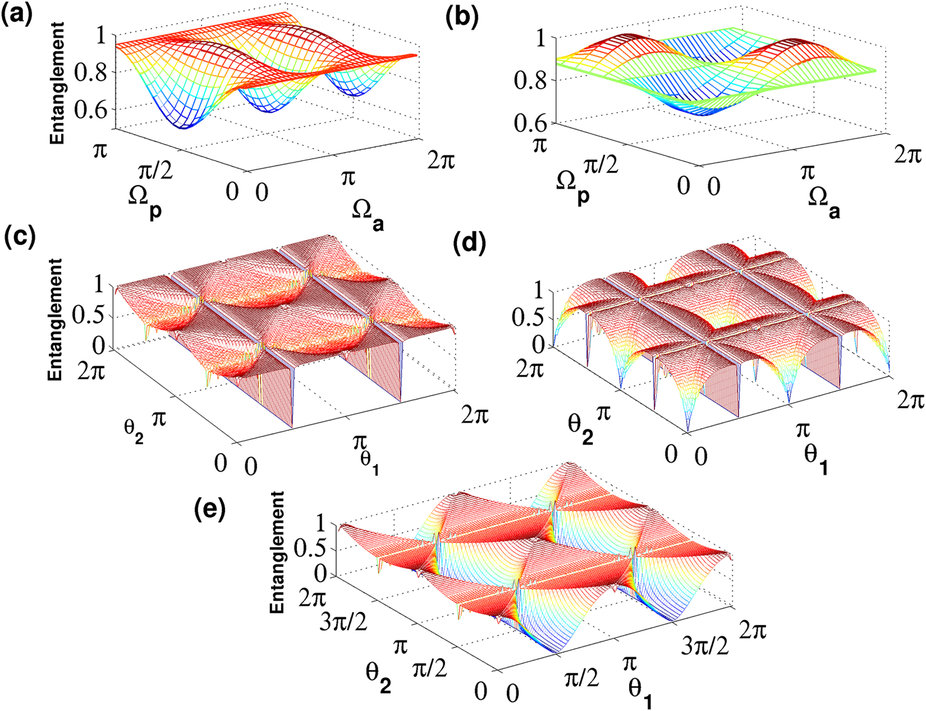} 
\caption{(color online) {\bf Entanglement between space and internal degree of freedom as a function of initial state and coin parameters after 90 steps of QW.} The entanglement as a function of initial state parameter for (a) split-step QW with $\theta_1 = 0 $, $\theta_2 = \frac{\pi}{4}$
and for (b) conventional QW with $\theta=\frac{\pi}{4}$ is shown. For (a) and (b) the azimuthal $\Omega_a$ and polar $\Omega_p$ angle correspond to the spherical coordinate angles of Bloch sphere associated with the internal degree (coin space). Entanglement as a function of $\theta_1$ and $\theta_2$ for initial state of the system (c) $\frac{1}{\sqrt{2}}(|\uparrow\rangle +i|\downarrow\rangle) \otimes |x=0\rangle$ (d) $|\uparrow\rangle \otimes |x=0\rangle$ and (e)$\frac{1}{\sqrt{2}}(|\uparrow\rangle + |\downarrow\rangle) \otimes |x=0\rangle$ is shown. The values of entanglement generated in split-step QW shows a significant dependency on the initial state.}
\label{fig4}
\end{figure}
In Fig.~\ref{fig4}(a) and (b), we show the profile of the entanglement as a function of azimuthal $\Omega_a$ and polar $\Omega_p$ angle of the initial state for split-step QW and conventional QW. From the plots we can observe that the range of the maximum and the minimum values is higher for split-step QW compared to the conventional QW.

In Fig. \ref{fig4}(c), (d) and (e) we show the value of entanglement as a function of parameter $\theta_1$ and $\theta_2$ for split-step QW with three different initial state, $\frac{1}{\sqrt 2}(|\uparrow\rangle + i |\downarrow \rangle) \otimes|x=0\rangle$, $|\uparrow\rangle\otimes|x=0\rangle$ and $\frac{1}{\sqrt 2}(|\uparrow\rangle + |\downarrow \rangle) \otimes|x=0\rangle$.

With two initial state parameters ($\Omega_p$ and $\Omega_a$) and the two coin operation parameters ($\theta_1$ and $\theta_2$), split-step QW will give more degrees of freedom to configure the dynamics resulting in maximum entanglement. For a constrained initial state, we can choose the evolution parameters to maximize the entanglement and for a constrained evolution parameters like the one leading to DCA, we can choose the initial state to maximize the entanglement.

\vskip 0.2in
\bc
{\bf Concluding Remarks}
\ec 

In summery, we have shown the recovery of the DCA and DH starting from the split-step QW. Earlier studies showed the gap in the connection between DCA and conventional QW due to the presence of the component in DCA which forced the probability amplitude to stay in the original position. 
Split-step QW which was developed to demonstrate greater control over the walk and explore topological phases by construction itself ensured the presence of probability amplitude at the original position during each step evolution. Exploiting this common feature in split-step QW and DCA, we analytically arrived at the combination of the two coin parameters $\theta_1$ and $\theta_2$ used in defining the split-step QW to recover the DCA and DH with all the fine oscillations in the probability distribution. In this work we have shown that the construction of split-step QW itself meet all the conditions required to arrive at DCA and DH without explicitly invoking any invariance condition. A similar equivalence relation between classical random walk and classical cellular automata is not known. Its the unitarity condition in quantum case that lead to this equivalence.

In our study, we also derived the expression for the Zitterbewegung oscillation from the parameters that define the split-step QW. This allows for identifying the parameters resulting in higher oscillations and its correspondence with real physical situation in the elementary particle dynamics where Zitterbewegung oscillations is observed. Variation of entanglement as a function of initial state and evolution parameters give greater degree of freedom to optimize the split-step QW for maximum entanglement compared to conventional QW. 
This simple connection between QW-DCA-DH could lead to an interesting regime of simulating free quantum field theory from the perspective of quantum information theory.  With quantum walk being used to simulate dynamics in various physical systems, it can soon play a very prominent role in designing a universal quantum simulator to simulate dynamics observed in both, condensed matter systems and quantum field theory.

\bc
\Large{\bf Methods}
\ec

\noindent
{\bf Recovery of Dirac Hamiltonian from Split-step QW~:~}  
In the equation~(\ref{UNSDQW}) if we put $\phi_1 = \phi_2 = \delta_1 = \delta_2 = 0 $ with a non-zero $\theta_1$ and $\theta_2$ values, 
we have the expression for the evolution operator,  
        \begin{align}
          U_{SQW} = \left( \begin{array}{cc}
              -\sin\theta_1 \sin\theta_2 ~ I     &          -i \cos\theta_1 \sin\theta_2 ~I \\
              + \cos\theta_1 \cos\theta_2 ~T_-   &          - i \sin\theta_1 \cos\theta_2 ~T_-\\ \\
             -i \cos\theta_1 \sin\theta_2 ~I    &          - \sin\theta_1 \sin\theta_2 ~I  \\
            - i \sin\theta_1 \cos\theta_2 ~T_+   &           + \cos\theta_1 \cos\theta_2 ~T_+\\
        \end{array}\right) = \left( \begin{array}{cc}
              -\sin\theta_1 \sin\theta_2  ~I  &          -i \cos\theta_1 \sin\theta_2~ I \\
              + \cos\theta_1 \cos\theta_2 ~e^{\frac{ipa}{\hbar}}  &  - i \sin\theta_1 \cos\theta_2 ~e^{\frac{ipa}{\hbar}}\\ \\
        -i \cos\theta_1 \sin\theta_2 ~I  & - \sin\theta_1 \sin\theta_2 ~I  \\
        - i \sin\theta_1 \cos\theta_2 ~e^{-\frac{ipa}{\hbar}}    & + \cos\theta_1 \cos\theta_2 ~e^{-\frac{ipa}{\hbar}}
        \end{array}\right) \label{actualSO}\end{align}
        
        This is an operator for both, on position space and internal degrees of freedom.
        
        Here we have used  the operator form \begin{align} e^{\pm  \frac{i p a}{\hbar} } = T_\mp, \nonumber\end{align} 
        with $p$ being the momentum operator, such that $ p\ket{k} = k \ket{k} $, where $\ket{k}$ is a momentum eigenbasis with momentum eigenvalue $k$.
  The above operator $U_{SQW}$ is diagonal in this momentum basis. So, from now onwards we will work with,     
       \begin{align} \bra{k}U_{SQW}\ket{k} =  U_{SQW}(k) = \left( \begin{array}{cc}
              -\sin(\theta_1) \sin(\theta_2)   &          -i \cos(\theta_1) \sin(\theta_2)~ \\
              + \cos(\theta_1) \cos(\theta_2) ~e^{\frac{ika}{\hbar}}  &  - i \sin(\theta_1) \cos(\theta_2) ~e^{\frac{ika}{\hbar}}\\ \\
        -i \cos(\theta_1) \sin(\theta_2)  & - \sin(\theta_1) \sin(\theta_2)  \\
        - i \sin(\theta_1) \cos(\theta_2) ~e^{-\frac{ika}{\hbar}}    & + \cos(\theta_1) \cos(\theta_2) ~e^{-\frac{ika}{\hbar}}
        \end{array}\right)\end{align} 
         is an operator on internal degrees of freedom only.  This operator is unitary,
         so this is a normal operator, hence diagonalizable. 
         Eigenvalues of operator $U_{SQW}(k) $ are,      
         
                            \begin{align} \Lambda_{\pm} = 
                                         \Big[ \cos(\theta_1) \cos(\theta_2) \cos\Big( \frac{ka}{\hbar} \Big)- \sin (\theta_1) \sin (\theta_2) \Big]
                                         \pm i \sqrt{ 1 -\Big[\cos (\theta_1) \cos (\theta_2) \cos \Big( \frac{ka}{\hbar} \Big) 
                                         - \sin (\theta_1) \sin (\theta_2) \Big]^2 }\nonumber\\
          = exp\Bigg( \pm i \cos^{-1}\Big[ \cos (\theta_1) \cos (\theta_2) \cos\Big( \frac{ka}{\hbar} \Big)- \sin (\theta_1) \sin (\theta_2) \Big]\Bigg)
                                         \end{align} with the definition,
                                         $ \cos(\frac{ka}{\hbar}) = \frac{1}{2} ( e^{\frac{ika}{\hbar}} + e^{-\frac{ika}{\hbar}} ) $.   
       
       The unnormalized eigenvectors of $ U_{SQW}(k) $ are, \begin{align}
                                                        \Big( \cos(\theta_1) \sin(\theta_2)  
                                                        + \sin (\theta_1) \cos (\theta_2) e^{\frac{i k a}{\hbar} } \Big) \ket{\uparrow} \nonumber\\ 
                                              + \Bigg( \cos (\theta_1) \cos (\theta_2) \sin\Big(\frac{ka}{\hbar}\Big)
                                              \mp \sqrt{ 1 -\Big[\cos (\theta_1) \cos (\theta_2) \cos\Big( \frac{ka}{\hbar}\Big)
                                              - \sin (\theta_1) \sin (\theta_2)  \Big]^2 }\Bigg)\ket{\downarrow} 
                                             \end{align}
                                             
     Their normalized eigenvectors of $U_{SQW}$ will be,
     \begin{align}
                                     \ket{\phi^\pm_k} = 
 N_k^\pm  \Bigg[ \Big( \cos(\theta_1) \sin(\theta_2)  + \sin (\theta_1) \cos (\theta_2) e^{\frac{i k a}{\hbar} } \Big) \ket{\uparrow} \otimes \ket{k} \nonumber\\ 
                                              + \Bigg( \cos (\theta_1) \cos (\theta_2) \sin\Big(\frac{ka}{\hbar}\Big)
                                              \mp \sqrt{ 1 -\Big[\cos (\theta_1) \cos (\theta_2) \cos\Big( \frac{ka}{\hbar}\Big) 
                                              - \sin (\theta_1) \sin (\theta_2)  \Big]^2 }\Bigg)\ket{\downarrow} \otimes \ket{k} \Bigg] \end{align}
                                        
              \begin{align}    
 N_k^\pm = \Bigg( 1 + \cos^2 (\theta_1) \sin^2(\theta_2) - \sin^2(\theta_1) \cos (2\theta_2) + \sin (2\theta_1) \sin (2\theta_2) \cos\Big(\frac{k a}{\hbar}\Big)\nonumber\\- 
\cos^2(\theta_1) \cos^2 (\theta_2) \cos \Big(\frac{2ka}{\hbar} \Big) 
\mp 2 \cos(\theta_1) \cos(\theta_2) \sin \Big( \frac{ka}{\hbar} \Big) 
\sqrt{1 - (\cos(\theta_1) \cos(\theta_2) \cos\Big(\frac{ka}{\hbar}\Big) - \sin(\theta_1) \sin(\theta_2))^2}\Bigg)^{-\frac{1}{2}}                           
\label{N}
\end{align}

     Now,  if unit time step evolution operator,  $ U_{SQW} := exp \big(-\frac{i H_{SQW} \tau}{\hbar} \big) $, 
        we will call ~$ H_{SQW} $,~ our effective Hamiltonian, which is also diagonal in momentum basis.
        So, in a similar manner, $U_{SQW}(k) := exp \big(-\frac{i H_{SQW}(k) \tau}{\hbar} \big) $,  
        then operator  $H_{SQW}(k) = \frac{i \hbar}{\tau} \ln [ U_{SQW}(k)] 
        = \frac{i \hbar}{\tau}\big[~(\ln\Lambda_+) \ket{\phi^+_k}\bra{\phi^+_k} ~+ ~(\ln\Lambda_-) \ket{\phi^-_k}\bra{\phi^-_k}~ \big].$

       Instead of this, in general we could 
        take,  \begin{align}H_{SQW}(k) = \frac{i \hbar}{\tau} \log_e [ U_{SQW}(k)]
        = \frac{i \hbar}{\tau}\big[~(\log_e\Lambda_+) \ket{\phi^+_k}\bra{\phi^+_k} ~+ ~(\log_e\Lambda_-) \ket{\phi^-_k}\bra{\phi^-_k}~ \big]\nonumber\\
        = \frac{i \hbar}{\tau}\big[~(\ln\Lambda_+ + 2\pi i r_+) \ket{\phi^+_k}\bra{\phi^+_k} ~
        + ~(\ln\Lambda_- + 2 \pi i r_-) \ket{\phi^-_k}\bra{\phi^-_k}~ \big],\end{align} where $r_+, r_-$ are integers. 
        Therefore, in general there would be an ambiguity of eigenvalue of $H_{SQW}(k)$ by an additional
        factor, $\frac{2 \pi \hbar}{\tau} \times$( an integer). 
        But in our formalism,  unitary evolution is fundamental and Hamiltonian is derived from this operator.
      We only need to see the effect of the Hamiltonian, in our evolution.
      So, we will consider $\frac{i \hbar}{\tau}\ln\Lambda_\pm $ mod $\frac{2 \pi \hbar}{\tau}$ as our energy eigenvalues,
      without further mentioning this `mod' operation.

        \begin{align} \frac{ i \hbar}{\tau} \ln \Lambda_{\pm} = 
        \mp \frac{\hbar}{\tau} \cos^{-1} \Bigg( \cos \theta_1 \cos \theta_2 \cos\Big( \frac{ka}{\hbar} \Big) 
        - \sin \theta_1 \sin \theta_2 \Bigg)  = \mp \hbar \omega_k \label{LAMBDA}. 
        \end{align}

       The eigenvectors of $U_{SQW}(k)$ are also eigenvectors of $ H_{SQW}(k) .$ We can form a unitary operator $V ,$ from the eigenvectors
                                             which diagonalizes $U_{SQW}(k)$ , that is,
                                             \begin{align} U_{SQW}(k) = V \left( \begin{array}{cc} 
                                                                                       \Lambda_+  & 0 \\
                                                                                       0 & \Lambda_- \\
                                                                                      \end{array} \right) V^{-1}.  \end{align}
Therefore,                                                                          
                                                                                     \begin{align} H_{SQW}(k) = \frac{ i \hbar}{\tau} V \left( \begin{array}{cc} 
                                                                                        \ln \Lambda_+  & 0 \\
                                                                                       0 &  \ln \Lambda_- \\
                                                                                      \end{array} \right) V^{-1} ~~
                                                                                      \Rightarrow ~~  H_{SQW}(k) =  \frac{i \hbar}{\tau} \ln (\Lambda_+ ) V \left( \begin{array}{cc} 
                                                                                       1  & 0 \\
                                                                                       0 & -1 \\
                                                                                      \end{array} \right) V^{-1}. 
                                                                                      \label{HSQW}
             \end{align}                                                                         
                                             
           In order to find out the mass term,  we put $ k = 0$ in the above eigenvalue equation (\ref{LAMBDA}),  we get the magnitude of mass 
             of the particle is equal to $|\theta_1 + \theta_2|$,  more correctly $|\theta_1 + \theta_2| + 2 \pi n,$  where n is an integer.

 From the equations~(\ref{LAMBDA}) and (\ref{HSQW}),
  we get the operator form of Hamiltonian operator in $\ket{k}$ basis,  \begin{align}
H_{SQW}(k) =                                                                                            \frac{ i \hbar \ln \Lambda_+}{ \tau \sqrt{1 -\Big(\cos (\theta_1) \cos (\theta_2) \cos\Big(\frac{ka}{\hbar}\Big) - \sin (\theta_1) \sin (\theta_2)  \Big)^2} }
                                                                                           \Bigg[\bigg(\cos(\theta_1) \cos (\theta_2) \sin\Big( \frac{ka}{\hbar}\Big) \bigg)
                                                                                                                      \left(\begin{array}{cc}
                                                                                                                                     1 & 0 \\
                                                                                                                                     0  & -1 \\
                                                                                                                                    \end{array}\right) \nonumber\\
                                                                                                                                    -  ( \cos(\theta_1) \sin (\theta_2) + \sin (\theta_1) \cos (\theta_2) e^{\frac{ika}{\hbar}}) 
                                                                                                                                                 \left(\begin{array}{cc}
                                                                                                                                                       0 & 1\\
                                                                                                                                                       0 & 0 \\
                                                                                                                                                      \end{array}\right) 
                                                                                                                                               -  ( \cos(\theta_1) \sin(\theta_2) + \sin(\theta_1) \cos(\theta_2) e^{-\frac{ika}{\hbar}} ) 
                                                                                                                                              \left( \begin{array}{cc}
                                                                                                                                              0 & 0 \\
                                                                                                                                              1 & 0 \\
                                                                                                                                             \end{array}\right) \Bigg].
                                                                                   \label{opeham}       
                                                                                   \end{align}
  This is the {\bf Hamiltonian operator for general split-step QW} in momentum basis.

    In the above equation (\ref{actualSO}) if we put , $ \sin(\theta_1) = 0 $, 
then our state evolution operator will be, 
\begin{align}
 U_{SQW} (\theta_1 = 0) = \sum_{x \in a\mathbb{Z}}  -i  \sin(\theta_2)  \Big[ \ket{x} \bra{x} \otimes \sigma_x \Big] 
 +  \cos (\theta_2)  \Big[  \ket{x - a} \bra{x} \otimes \ket{\uparrow} \bra{\uparrow} 
 +   \ket{x + a} \bra{x}  \otimes \ket{\downarrow}\bra{\downarrow} \Big] \equiv U_{DA}.
 \end{align}
 
    Then, $\alpha = \cos (\theta_2) $ and $\beta = \sin (\theta_2) $  with the constraint $ \sin (\theta_2) \in [0,1] $.

   Also the  Hamiltonian (equation~(\ref{opeham})) boils down to the Hamiltonian for Dirac cellular automata, in momentum basis, for $\sin\theta_1 = 0, $
   \begin{align}
    H_{SQW}(k, \theta_1 = 0) 
    = - \frac{ \hbar \cos^{-1} (\cos (\theta_2) \cos (k))}{ \tau \sqrt{1 - (\cos (\theta_2) \cos k)^2}} \Bigg[ \cos(\theta_2) \sin k \left( \begin{array}{cc}
                                                                                                                       1 & 0\\
                                                                                                                       0 & -1 \\
                                                                                                                      \end{array}\right) -\sin \theta_2 \left( \begin{array}{cc}
                                                                                                                                                               0 & 1\\
                                                                                                                                                               1 & 0 \\
                                                                                                                                                              \end{array}\right)
 \Bigg] \equiv H_{DA}
   \end{align}
We previously identified,  in the general  split-step QW case,  mass = $|\theta_1 + \theta_2|,$ so in this case mass is equals $\theta_2,$ as $\theta_1 = 0$. 

And for smaller mass $\theta_2  \approx 0,$ momentum $ k \approx 0,$ $\sin\theta_2 \approx \theta_2, \cos\theta_2 \approx 1, \sin\Big(\frac{ka}{\hbar}\Big) \approx \frac{ka}{\hbar}, \cos\Big(\frac{ka}{\hbar}\Big) \approx 1. $
Then \begin{align}
      H_{DA} \approx -\frac{a}{\tau} k \left(\begin{array}{cc}
                                1 & 0\\
                                0 & -1\\
                               \end{array}\right) + \frac{\hbar}{\tau} \theta_2 \left( \begin{array}{cc}
                                                                 0 & 1\\
                                                                 1 & 0\\
                                                                 \end{array}\right)
                               \end{align} which is in a form of one-dimensional Dirac Hamiltonian for a $\frac{1}{2} $ spinor.

{\bf Derivation of Zitterbuguang frequency~:~}
For the case of general split-step QW, the state $\ket{\chi} $ of a particle moving with momentum $k$,
 can be expressed as a linear superposition of the energy eigenstates $\ket{\phi^\pm_k}$ (normalized) with the same momentum $k$ , so 
 \begin{align}
                       \ket{\chi(t=0)} = c_1 \ket{\phi^+_k(t=0)} + c_2 \ket{\phi^-_k(t=0)} \nonumber\\
                        \Rightarrow~~ \ket{\chi(t)} = c_1 ~ (U_{SQW})^\frac{t}{\tau} ~~ \ket{\phi^+_k(t=0)} ~ + ~ c_2 ~ (U_{SQW})^\frac{t}{\tau}~~ \ket{\phi^-_k(t=0)} \nonumber\\
                       \Rightarrow~~ \ket{\chi(t)} = c_1 e^{-\frac{i H t}{\hbar} }~~ \ket{\phi^+_k(t=0)} ~+~ c_2 e^{-\frac{iHt}{\hbar}}~~ \ket{\phi^-_k(t=0)} \nonumber\\
                       \Rightarrow ~~ \ket{\chi(t)} = c_1 e^{i \omega_k t }~~ \ket{\phi^+_k(t=0)} ~+~ c_2 e^{-i \omega_k t}~~ \ket{\phi^-_k(t=0)} \nonumber\\
                        \Rightarrow ~~ < \hat{A}>_t =  \bra{\chi(t)} \hat{A} \ket{\chi(t)}             
                        = |c_1|^2 \bra{\phi^+_k(t=0)} \hat{A} \ket{\phi^+_k(t=0)} + |c_2|^2 \bra{\phi^-_k(t=0)} \hat{A} \ket{\phi^-_k(t=0)} \nonumber\\
                        +   c^*_1 c_2 e^{ - 2 i \omega_k t} \bra{\phi^+_k(t=0)} \hat{A} \ket{\phi^-_k(t=0)} +  c_1 c^*_2 e^{  2 i \omega_k t} \bra{\phi^-_k(t=0)} \hat{A} \ket{\phi^+_k(t=0)}.
\label{eqpecval}                                                         
                                                                          \end{align}
 
Here $|c_1|^2 + |c_2|^2 =1$ where $c_1$ and $c_2$ are complex numbers. 
In the equation~(\ref{eqpecval}), $t$ is a integer multiple of $\tau$,
if we do not consider this, then we have to take,  integral part of $\Big( \frac{t}{\tau} \Big)$ instead of $\frac{t}{\tau}.$ We can see that the time dependent part,
\begin{align}
  c^*_1 c_2 e^{  -2 i \omega_k t} \bra{\phi^+_k(t=0)} \hat{A} \ket{\phi^-_k(t=0)} +  c_1 c^*_2 e^{  2 i \omega_k t}
  \bra{\phi^-_k(t=0)} \hat{A} \ket{\phi^+_k(t=0)} \nonumber\\
= \{2 \Re(c^*_1 c_2 \bra{\phi^+_k(t=0)} \hat{A} \ket{\phi^-_k(t=0)})\} ~\cos(2 \omega_k t) + 
\{ 2 \Im(c^*_1 c_2 \bra{\phi^+_k(t=0)} \hat{A} \ket{\phi^-_k(t=0)}) \} ~ \sin(2 \omega_k t).
  \end{align}
It contains frequency,
\begin{align} \frac{2 \omega_k}{2 \pi} = \frac{\bra{k} H_{SQW} \ket{k}}{ \pi \hbar}.\nonumber\end{align} which is identified as 
   the Zitterbewegung frequency,  
            \begin{align}
           Z_{SQW} =  \frac{ 1 }{\tau \pi} \cos^{-1} \Big[ \cos \theta_1 \cos \theta_2 \cos\Big(\frac{ka}{\hbar}\Big) - \sin \theta_1 \sin \theta_2 \Big]
            \end{align}
        and corresponding amplitude of oscillation = 
        \begin{align}
         \mathcal{A}_{SQW} = 2 \sqrt{  \big[\Re(c^*_1 c_2 \bra{\phi^+_k(t=0)} \hat{A} \ket{\phi^-_k(t=0)})\big]^2 +
         \big[ \Im(c^*_1 c_2 \bra{\phi^+_k(t=0)} \hat{A} \ket{\phi^-_k(t=0)})\big]^2},
       \end{align}
  where,  \begin{align}
          \bra{\phi^+_k(t=0)} \hat{A} \ket{\phi^-_k(t=0)} \nonumber\\
          = N^+_k N^-_k \Bigg[ \big\{ \cos^2 \theta_1 \sin^2 \theta_2 
          + \sin^2 \theta_1 \cos^2 \theta_2
          + 2 \sin \theta_1 \sin \theta_2 \cos \theta_1 \cos \theta_2  
          \cos\Big(\frac{ka}{\hbar} \Big)\big\} \bra{\uparrow}\bra{k} \hat{A} \ket{k}\ket{\uparrow}\nonumber\\
          -  \big\{ \cos^2\theta_1 \sin^2 \theta_2  + \sin^2 \theta_1 \cos^2 \theta_2
          + 2 \sin\theta_1  \sin \theta_2 \cos \theta_1 \cos\theta_2 \cos \Big(\frac{ka}{\hbar}\Big)\big\}
          \bra{\downarrow}\bra{k} \hat{A} \ket{k}\ket{\downarrow}\nonumber\\ 
          + \big\{ (\cos\theta_1 \sin\theta_2 + \sin\theta_1 \cos \theta_2 
          e^{-\frac{ika}{\hbar}})(\cos\theta_1 \cos\theta_2 \sin\Big(\frac{ka}{\hbar}\Big) +  \sin(\pi \tau Z_{SQW} ) )\big\}
                    \bra{\uparrow}\bra{k} \hat{A} \ket{k}\ket{\downarrow}\nonumber\\
                   + \big\{ (\cos\theta_1 \sin\theta_2 + \sin\theta_1 \cos \theta_2 
          e^{\frac{ika}{\hbar}})(\cos\theta_1 \cos\theta_2 \sin\Big(\frac{ka}{\hbar}\Big) - \sin(\pi \tau Z_{SQW} ) )\big\}
                    \bra{\downarrow}\bra{k} \hat{A} \ket{k}\ket{\uparrow} \Bigg].
         \end{align}
         $N^\pm_k$ is given by equation (\ref{N}).

\vskip 0.2in
\noindent
{\bf Acknowledgment:}

\noindent
CMC would like to thank A. P\'erez for useful discussions and Department of Science and Technology, Government of India for the Ramanujan Fellowship grant No.:SB/S2/RJN-192/2014. 

\vskip 0.2in

\noindent  
{\bf Author Contributions:}

\noindent
CMC designed the study, AM and CMC together carried out analytical derivation, numerical analysis, prepared figures, interpreted the results, and wrote the manuscript.

\vskip 0.2in
\noindent
{\bf Competing financial interests :}

\noindent
The author declare no competing financial interests.



\begin{thebibliography}{99}



\bibitem{Yuk66} Yukawa, H. Atomistics and the Divisibility of Space and Time.
{\it Prog. Theor. Phys. Suppl. {\bf 37} and {\bf 38}, 512 (1966).}

\bibitem{Yam84} Yamamoto, H. Quantum field theory on discrete space-time.
 {\it Phys. Rev. D {\bf 30}, 1127 (1984)}.

\bibitem{Wil74} Wilson, K. G. Confinement of quarks 
{\it Phys. Rev. D {\bf 10}, 2445 (1974)}.

\bibitem{BMS85} Bender, C. M., Milton, K. A. \& Sharp, D. H. Gauge invariance and the finite-element solution of the Schwinger model 
{\it Phys. Rev. D {\bf 31}, 383 (1985).}

\bibitem{Bia94}Bialynicki-Birula, I. Weyl, Dirac, and Maxwell equations on a lattice as unitary cellular automata. 
{\it Phys. Rev. D {\bf 49}, 6920 (1994)}.

\bibitem{Kog79} Kogut, J. B. An introduction to lattice gauge theory and spin systems. 
{\it Rev. Mod. Phys. {\bf 51}, 659 (1979)}.

\bibitem{Rot05} Rothe, H. J. Lattice gauge theories: An Introduction.
{\it World Scientific Lecture Notes in Physics {\bf 74}. World Scientific, Singapore. (2005)}.

\bibitem{Neu66} Neumann, J. von.  {\it Theory of Self-Reproducing Automata, University of Illinois Press, Urbana, London, 1966}.

\bibitem{W09} Wiesner,K., Quantum Cellular Automata.
{\it Encyclopedia of Complexity and Systems Science, pp 7154-7164 (Springer New York, 2009)}.

\bibitem{Mey96} Meyer, D. A. From quantum cellular automata to quantum lattice gases. 
{\it J. Stat. Phys. {\bf 85},  551 (1996)}.


\bibitem{Bisio}  Bisio, A.,   D’Ariano,G. M.,  Tosini, A. 
Quantum field as a quantum cellular automaton: The Dirac free evolution in one dimension
{ \it Annals of Physics   {\bf 354}, 244–264 (2015)}.

\bibitem{DP14} D\`Ariano, G. M., \&  Perinotti, P. Derivation of the Dirac equation from principles of information processing.
{\it Phys. Rev. A {\bf 90}, 062106 (2014).}

\bibitem{Joc95} Jacobson, T., Thermodynamics of Spacetime: The Einstein Equation of State
{\it Phys. Rev. Lett. {\bf 75} 1260 (1995).}

\bibitem{Ver11} Verlinde, E. On the origin of gravity and the laws of Newton. 
{\it J. High Energy Phys. {\bf 04}, 029 (2O11).}

\bibitem{Ria58} Riazanov,  G. V. The Feynman path integral for the Dirac equation. {\it Sov. Phys. JETP {\bf 6}, 1107-1113 (1958)}.

\bibitem{Fey86} Feynman, R. P. Quantum mechanical computers.
{\it Found. Phys. {\bf 16}, 507-531 (1986)}.

\bibitem{Par88} Parthasarathy, K. R. The passage from random walk to diffusion in quantum probability.
{\it Journal of Applied Probability, {\bf 25}, 151-166 (1988)}.

\bibitem {ADZ93} Aharonov, Y.,  Davidovich, L. \& Zagury,  N. Quantum random walks. 
{\it Phys. Rev. A {\bf 48}, 1687-1690 (1993)}.

\bibitem{HKS05}  Hamada,M.,  Konno, N., Segawa, E.  Relation between coined quantum walks and quantum cellular automata. 
{\it RIMS Kokyuroku, {\bf No.1422}, pp.1-11 (2005)}.



\bibitem{asymp} A. P\'erez,  Asymptotic properties of the Dirac quantum cellular automaton.
 {\it Phys. Rev. A {\bf 93}, 012328 (2016)}.
 


\bibitem{Sal12} Venegas-Andraca,  S. E. Quantum walks: a comprehensive review. 
{\it Quantum Information Processing  {\bf 11} (5), pp. 1015-1106 (2012)}.


\bibitem{HM09} Hoyer, S. \& Meyer, D. A. Faster transport with a directed quantum walk.
{\it Phys. Rev. A {\bf 79}, 024307 (2009)}.

\bibitem{PH08}  Plenio, M. B. \& Huelga, S. F.  Dephasing-assisted transport: quantum networks and biomolecules.
{\it New J. Phys. {\bf 10}, 113019 (2008)}.

\bibitem{CB15} Chandrashekar, C. M. \& Busch, Th. Localized quantum walks as secured quantum memory.
{\it European Physics Letters 110, 10005 (2015).}

\bibitem{ECR07} Engel, G. S. {\it et al.}, Evidence for wavelike energy transfer through quantum coherence in photosynthetic systems. 
{\it Nature {\bf 446}, 782-786 (2007)}.  

\bibitem{MRL08} Mohseni, M., Rebentrost, P., Lloyd, S. \& Aspuru-Guzik, A. Environment-assisted quantum walks in photosynthetic energy transfer. 
{\it J. Chem. Phys. {\bf 129}, 174106 (2008)}. 





\bibitem{KFC09} Karski, K. {\it et al.} Quantum walk in position space with single optically trapped atoms. 
{\it Science  {\bf 325}, 174 (2009)}.


\bibitem{PLM10} Peruzzo, A. {\it et al.} Quantum walks of correlated photons. 
{\it Science {\bf 329}, 1500 (2010)}.

\bibitem{SCP10} Schreiber, A. {\it et al.} Photons walking the line: a quantum walk with adjustable coin operations.
{\it Phys. Rev. Lett. {\bf 104}, 050502 (2010)}.

\bibitem{BFL10} Broome, M. A.  {\it et al.} Discrete single-photon quantum walks with tunable decoherence.
{\it Phys. Rev. Lett. {\bf 104}, 153602 (2010)}.



\bibitem{LMT10} L. Lepori, G. Mussardo, \& A. Trombettoni, (3 + 1) massive Dirac fermions with ultracold atoms in frustrated cubic optical lattices.
{\it Europhys. Lett. {\bf 92}, 50003 (2010)}.

\bibitem{BMR10}  Bermudez,A. ,  Mazza, L.,  Rizzi, M.,  Goldman, N. , Lewenstein, M. \&  M. Martin-Delgado,
Wilson Fermions and Axion Electrodynamics in Optical Lattices.
{ \it Phys. Rev. Lett. {\bf 105}, 190404 (2010)} .

\bibitem{CMP10}  Cirac,J. I. , Maraner, P. \&   Pachos,J.
Cold Atom Simulation of Interacting Relativistic Quantum Field Theories.
{ \it Phys. Rev. Lett. {\bf 105}, 190403 (2010)}.

\bibitem{SP11}  Semiao, F. L. \&  Paternostro, M. 
Quantum circuits for spin and flavor degrees of freedom of quarks forming nucleons. {\it Quant. Inf. Proc. {\bf 11}, 67–75 (2011).}

\bibitem{KM11} Kapit, E \&  Mueller, E. Optical-lattice Hamiltonians for relativistic quantum electrodynamics.
{\it Phys. Rev. A {\bf 83}, 033625 (2011)}.


\bibitem{HBP06}Hartmann, M. J.,   Brandao, F. G. S. L.  \&   Plenio, M. B. Strongly interacting polaritons in coupled arrays of cavities.
{ \it Nat. Phys. {\bf 2}, 849 (2006).}

\bibitem{GTC06} Greentree, A. D.,   Tahan,C. ,   Cole, J. H., \&    Hollenberg,L. C. L. Quantum phase transitions of light.
{ \it Nat. Phys. {\bf 2}, 856 (2006).}

\bibitem{ASB07}  Angelakis, D.G.,  Santos, M. F., \&  Bose, S. 
Photon-blockade-induced Mott transitions and XY spin models in coupled cavity arrays.
{\it  Phys. Rev. A {\bf 76}, 031805 (2007)}.

\bibitem{BR12} Blatt, R. \& Roos, C. F. Quantum simulations with trapped ions.
{ \it Nature Phys. {\bf 8}, 277-284 (2012)}.

\bibitem{AW12} Aspuru-Guzik, A. \& Walther, P. Photonic quantum simulators.  {\it Nature Phys. {\bf 8}, 285-291 (2012).} 



\bibitem{NC00} Nielsen, M. A. \& Chuang, I. L. {\it Quantum Computation and Quantum Information}.~  (Cambridge University Press, 2000).

\bibitem{Fey82} Feynman, R. Simulating physics with computers.
{\it Int. J. Theor. Phys. {\bf 21}, 467 (1982)}.

\bibitem{Whe90} Wheeler, J. A. Information, Physics, Quantum : The Search for Links. 
In {\it Complexity Complexity, Entropy, and the Physics of Information}, Ed W. H. Zurek (Redwood City, CA: Addison-Wesley) (1990).

\bibitem{JLP12} Jordan, S.P.,Lee,  K. S. M., \& Preskill, J.  Quantum Algorithms for quantum field theories. 
 {\it Science {\bf Vol. 336}, Issue 6085, pp. 1130-1133 (Jun 2012)}.











\bibitem{CBS10} Chandrashekar, C. M., Banerjee, S. \& Srikanth, R. Relationship between quantum walks and relativistic quantum mechanics.
{\it  Phys. Rev. A  {\bf 81}, 062340 (2010)}.

\bibitem{Fre06}Strauch, F. W. Relativistic quantum walks.
 {\it Phys. Rev. A {\bf 73}, 054302 (2006).}

\bibitem{Cha13} Chandrashekar, C. M. Two-component Dirac-like Hamiltonian for generating quantum walk on one-, two- and three-dimensional lattices. 
{\it Sci. Rep. {\bf 3}, 2829; DOI:10.1038/srep02829 (2013)}. 


\bibitem{MD12} Molfetta, G. D. \& Debbascha, F. Discrete-time quantum walks: Continuous limit and symmetries. 
{\it Journal of Mathematical Physics {\bf 53}, 123302 (2012)}.

\bibitem{MBD13} Molfetta, G. D., Brachet, M. \&  Debbasch, F. Quantum walks as massless Dirac fermions in curved space-time. 
 {\it Phys. Rev. A  {\bf 88}, 042301 (2013)}.

\bibitem{AFF15}Arrighi, P., Facchini, S. \& Forets, M. Quantum walks in curved spacetime. {\it arXiv:1505.07023 (2015)}



\bibitem{MBD14} Molfetta, G. D., Brachet, M. \&  Debbasch, F. Quantum walks in artificial electric and gravitational fields.
{\it Physica A {\bf 397}, 157–168 (2014)}. 


 
\bibitem{SFP15} Succi, S., Fillion-Gourdeau, F., \&  Palpacelli, S. Quantum lattice Boltzmann is a quantum walk. 
{EPJ Quantum Technology {\bf 2}, 12 (2015).}

\bibitem{kita}  Kitagawa, T.,  Rudner,M.S.,  Berg, E.  \&  Demler, E.  Exploring topological phases with quantum walks.
{\it Phys. Rev. A { \bf 82}, 033429 (2010)}.

\bibitem{kari}  Kari, J. Theory of cellular automata: A survey.
{\it Theoretical Computer Science 334, 3-33 (2005)}.

\bibitem{GZ88}  Grossing, G.,   Zeilinger, A.      Quantum Cellular Automata
{\it Complex Systems  {\bf 2}, 197-208 (1988)}.

\bibitem{Wat95} Watrous, J.  On One-Dimensional Quantum Cellular Automata.
{\it 36th Annual Symposium on Foundations of Computer Science, Proceedings, IEEE, pp. 528-537; DOI:10.1109/SFCS.1995.492583 (1995).}

\bibitem{zit}  D\'avid, G.,  Cserti, J. General theory of Zitterbewegung.
{\it Phys. Rev. B  {\bf 81}, 121417R  (2010)}.


\bibitem{Cha12} Chandrashekar, C.M. Disorder induced localization and enhancement of entanglement in one- and two-dimensional quantum walks.
{\it arXiv:1212.5984 (2012).}



 \end{thebibliography}
\end{document}